# zenDot: An LLM-integrated quantum TCAD platform for semiconductor quantum-device design and optimization automation

Zeheng Wang[1,2,*], Yan Liu[1,2], Yue Hao[1,2], and Genquan Han[1,2]

[1] Hangzhou Institute of Technology, Xidian University, Hangzhou 311231, China

[2] Faculty of Integrated Circuit, Xidian University, Xi'an 710071, China

Email: zenwang@outlook.com

## Abstract

Semiconductor quantum-device design still lacks an integrated Technology Computer-Aided Design (TCAD)-like environment that connects material geometry, quantum many-body simulation, and automated design. Here we introduce zenDot, a large-language model (LLM)-integrated quantum TCAD platform that links a material-labelled device state to a unified condensed-matter physics toolbox. The device and calculation components are integrated into a desktop workbench, Python API, and an embedded LLM agent, allowing electrostatics, charge and transport characterization, correlated-state calculations, and qubit modelling to be executed within one reproducible environment. We demonstrate zenDot on a $Si/SiO_2$ double quantum dot, where a single device state reproduces the characterization workflow and supports hybrid, tunnel-charge, and singlet-triplet qubit analyses. A platform-level universal-control scan revises the singlet-triplet operating point and reduces the predicted worst-gate infidelity by nearly 30-fold. Beyond analysis, the LLM agent directly operates the same physics environment as human users, proposing design changes, executing registered simulations, and iterating on solver-returned metrics under physics-aware validation. Across three demonstration tasks it completes 18 validated design iterations, including geometry modification followed by a full re-solve from the material stack. zenDot establishes a machine-operable quantum TCAD workflow that connects device physics with LLM-driven design exploration.

## Introduction

Technology computer-aided design (TCAD) transformed classical semiconductor development by turning materials, geometry, process conditions, and electrical performance into an iterative computational workflow before fabrication. Semiconductor quantum devices are reaching a similar design bottleneck. Their behaviour depends not only on electrostatics and transport, but also on discrete spectra, tunnel and exchange couplings, many-body states, coherence, initialization, and gate operations. These quantities span different physical descriptions and must be evaluated across a design space that includes material stacks, gate geometry, operating bias, charge configuration, and logical encoding.

Conventional device TCAD does not provide this complete route from a physical structure to quantum-device performance. [1-7,13-17]

The required physics is not missing; it is fragmented. Self-consistent quantum-dot electronic structure, charge-stability and transport simulation, correlated-state methods, and non-equilibrium quantum transport have each matured into powerful specialist approaches. In practice, however, they are often connected through separate codes, re-entered parameters, and study-specific scripts. This fragmentation makes it difficult to treat a quantum device as one continuously evolving design object. What is needed is therefore not another isolated solver, but an environment in which a material-defined device can move through electrostatics, characterization, many-body analysis, and qubit modelling without rebuilding the physical state at every stage. [8-12,18-31]

Large language models add a new reason to solve this integration problem. Tool-using LLMs can already plan multi-step scientific and engineering workflows, while automated methods increasingly assist quantum-dot tuning and charge-state recognition. A quantum-device design agent, however, is useful only when it can act on real simulation tools rather than merely describe what should be calculated. This requires a machine-operable physics environment in which the model can select calculations, change device or control parameters, launch solvers, receive quantitative results, and continue the design loop through the same interfaces available to a human researcher. [32-37]

Here we present zenDot, an LLM-integrated quantum TCAD platform built around this idea. zenDot combines a material-labelled device representation, a broad condensed-matter physics toolbox, and a common component registry that is exposed identically to the desktop workbench, Python API, and LLM agent. We first establish the platform architecture and its device-to-physics workflow. We then use a $Si/SiO_2$ double quantum dot to demonstrate characterization from the layer stack and to show that one calibrated device state can support three distinct qubit models. Finally, we place the LLM inside the same computational environment and demonstrate iterative design over both electrical controls and device geometry. The qubit calculations serve as demonstrations of platform breadth; the central result is a quantum-device environment that can be operated by both researchers and language models.

## zenDot platform

### One device state from geometry to quantum observables

zenDot starts from the physical device. A project script defines materials, layers, gates, reservoirs, and electrical controls, and the resulting device state is shared by every downstream calculation. Material-resolved electrostatics produces the channel confinement, from which localized dots and device-level quantities such as lever arms, capacitance terms, tunnel coupling, and exchange are derived. The same object is then passed to charge, transport, correlated-state, and encoded-qubit calculations. In this way, the platform treats geometry, electrostatics, and quantum observables as successive views of one device rather than as independent inputs assembled after the fact (Fig. 1a-d).

The present platform exposes 18 calculation components and 18 visualization components through one registry. The registry spans electrostatics, static and finite-bias charge maps, spectroscopy, correlated states, dynamics, quantum transport, and qubit operations. Each component declares the parameters and device capabilities it requires and the result type it produces. The desktop application builds its calculation panels from this registry, Python calls the same component identifiers, and the LLM agent invokes the same tools. A calculation selected by the agent is therefore the same calculation available to a researcher at the workbench or in a script, not a separate AI-specific approximation (Fig. 1e,f).

This shared computational path is central to the LLM integration. Numerical outputs carry their units, solver context, and device state so that subsequent calculations, figures, and optimization objectives are connected to the simulation that produced them. The implementation also records the validity range of each component, allowing a broad physics toolbox to coexist without turning unlike observables into interchangeable numbers. These details are transparent to ordinary use but provide the structure required for reproducible multi-stage design.

For automated operation, zenDot adds a physics-aware execution layer on top of the same registry. The LLM can propose a calculation or a bounded change to the declared design variable; zenDot checks that the action is complete and compatible with the current device, executes the registered solver, and returns the measured result to the design loop. This keeps the language model in the role where it is most useful - selecting and sequencing computational actions - while the physics engines remain responsible for the numerical evidence (Fig. 1g).

## A workbench for human and machine operation

The desktop workbench provides device construction, physical settings, component selection, map configuration, result inspection, export, documentation, and the integrated Assistant within one application (Fig. 2). Component availability follows the loaded device and selected backend, while visualization choices follow the result returned by each calculation. Saved projects preserve the device definition and simulation settings needed to reconstruct a study. The Assistant is therefore not an external chatbot attached to a simulator: it is another operating surface of the same quantum-device environment.

# Integrated quantum-physics workflow

## A condensed-matter toolbox behind one device model

Quantum-device TCAD requires a different physics stack from classical drift-diffusion simulation. zenDot therefore integrates calculations for equilibrium charge and single-electron behaviour, sequential and finite-bias transport, magnetospectroscopy, real-time and quasi-static dynamics, response functions, Wigner-molecule correlations, matrix-product ground states, and non-equilibrium Green-function transport. The transport layer includes surface-Green-function decimation, Fisher-Lee conversion, multiband models, nonorthogonal bases, density-functional NEGF, and time-dependent propagation. These methods are not presented as one universal approximation; they are a toolbox for different regimes of the same device-design problem. [6,10-12,18-30]

The value of combining these methods is the continuity of the design workflow. A researcher can move from a layer stack to confinement, charge-state structure, tunnelling, spectroscopy, correlated states, and qubit-level observables without rebuilding the device in a second software path. Components become available when their physical requirements are satisfied, and their outputs retain the model and parameter context needed by downstream calculations. Detailed numerical algorithms and approximation ranges are given in the Supporting Information.

## From the material stack to device characterization

We demonstrate this workflow on a lateral $Si/SiO_2$ double quantum dot. The two plunger gates lie on the lower z = 78 nm plane, while the source, inter-dot, and drain barrier gates lie on the upper z = 81 nm plane. The 3 nm vertical separation between the gate planes produces, under the chosen controls, two resolved confinement minima separated by 42 nm. Independent calculations of the left and right dots return the same lever arm of 0.28941 eV $V^{-1}$, charging energy of 36.73 meV, addition energy of 36.18 meV, and electron g factor of 2.000, consistent with the mirror symmetry of the reference device (Fig. 3a-f,k).

The same device state then generates the two-dot observables used for subsequent qubit calculations. The charge-stability map resolves four labelled cells including (0,0), with 139 mV of headroom below the first charging line and an interior (1,1) operating point at a common-mode plunger offset of 0.540 V (Fig. 3g-j). At the reference barrier, the exchange energy is 1.019 µeV and the tunnel coupling is 93.174 µeV. The mirror-symmetric device gives an even exchange function $J(\varepsilon)$; in this model $\varepsilon = 0$ is a cusp-shaped local maximum, so the sensitivity at the symmetry point is evaluated from the one-sided slope rather than from a centred difference across the cusp (Fig. 3l). Together, these calculations establish the device state inherited by the encoded-qubit demonstrations.

# One device state, three qubit models

## Cross-encoding demonstration of the platform

The reference double dot provides a compact test of whether one platform can support models that use different electron numbers, operating biases, and Hamiltonians. We therefore evaluate hybrid, tunnel-charge, and singlet-triplet qubits from the same device-defined electrostatic and coupling state. Each model receives the device quantities relevant to its Hamiltonian and returns its own spectrum, dynamics, coherence, thermal, and gate metrics. The purpose is not to rank the three encodings, but to show that a single device model can feed distinct qubit analyses without rebuilding the physical device for each calculation.

## Hybrid-qubit demonstration

For the three-electron hybrid qubit, the logical splitting approaches the valley splitting and reaches 348.104 μeV at the reported operating point, corresponding to 84.171 GHz. Within the retained charge-noise model and the ±60 mV detuning window, the calculation gives $T_2^*$ = 887.3 ns and a quality factor $Q$ = 149362 for a π rotation. The logical initialization fidelity is 1.000 at 100 mK because the logical splitting is approximately 40 $k_B T$. The complete sweep, thermal populations, and window dependence are shown in Fig. 4 and the Supporting Information.

## Tunnel-charge-qubit demonstration

The tunnel-charge qubit uses the device-derived inter-dot tunnel coupling as a two-level anticrossing. At zero detuning, $2t_c$ = 186.347 μeV, corresponding to 45.059 GHz. The same model gives $T_2^*$ = 4.697 ns, a modelled infidelity of 1.064 × $10^{-3}$, and $Q$ = 846.6 for a π/2 rotation. Zero detuning maximizes dephasing time in the retained charge-noise model, while the full gate calculation also captures the finite rotation-axis error. The logical initialization fidelity is 1.000 at 100 mK (Fig. 5).

## Singlet-triplet-qubit demonstration

The singlet-triplet calculation provides a more consequential design example. With a 25 mT dot-to-dot field difference, the rotation axis lies 70.6° from z and the worse gate of a universal pair has an infidelity of 0.759, even though the isolated z-gate error is 5.99 × $10^{-8}$. When zenDot scans the Zeeman gradient against the complete universal-control objective, it instead selects $\Delta E_Z$ = 0.220 μeV (1.90 mT). At this point the splitting is 1.043 μeV with $J$ = 1.019 μeV, the frequency is 252.2 MHz, $T_2^*$ = 2783 ns, and the predicted worst-gate infidelity falls to 0.0259 (Fig. 6).

This approximately 29-fold change in the worst-gate error illustrates the role of the platform rather than establishing a universal optimum for singlet-triplet qubits. The selected point trades charge-noise suppression against coherent axis error, and zenDot exposes both quantities within the same calculation. More broadly, the three encoding branches show that speed, coherence, initialization, and gate error can be evaluated from one device state while remaining tied to the Hamiltonian that defines each metric.

## LLM-driven design exploration

### The LLM operates the physics platform, not a surrogate

The final demonstration places the LLM inside the design loop. Each study gives the agent a design objective, one allowed variable, its physical bounds, and a simulation budget. The model then proposes the next action using the same registered components available to a human user. zenDot executes the requested calculation, extracts the objective from the solver result, returns the measured outcome, and makes that information available for the next proposal. The agent therefore interacts with a quantum-device simulator as an experimental design environment: it chooses what to change and what to calculate, while the numerical result comes from the underlying physics model (Fig. 7a).

Across three demonstration tasks, the LLM completes 18 solver-validated design iterations. In the hybrid-qubit task, six barrier proposals increase Q from 161412.67 to 220112.30, with the best sampled point at -0.300 V. In the tunnel-charge task, six proposals increase Q from 225.07 to 809.67, with the best point at -0.028906 V. These campaigns are intentionally small, but they demonstrate a complete closed loop from natural-language planning to parameter change, physics execution, quantitative feedback, and another model action (Fig. 7b-g).

The third task moves beyond tuning an electrical control. The LLM varies the inter-dot gate length, causing zenDot to modify the physical geometry and rebuild the device before evaluating the singlet-triplet objective. Six geometry-level iterations reduce the measured infidelity from $6.273816 \times 10^{-8}$ to $6.272891 \times 10^{-8}$ at a gate length of 22.5 nm (Fig. 7h-j). Although the numerical improvement in this particular landscape is small, the workflow is the important result: the language model is not optimizing a fitted response surface, but editing a device and triggering a fresh physics calculation from the material stack. Scanned landscapes and complete agent histories are retained in the Supporting Information for quantitative comparison of the sampled trajectories.

# Discussion

zenDot reframes semiconductor quantum-device simulation as a connected design environment rather than a collection of independent calculations. The platform starts from a material-defined device and carries that state through electrostatics, charge and transport characterization, many-body physics, and encoded-qubit analysis. The same computational components are accessible through a graphical workbench, Python, and an LLM interface. This combination is important because machine operation is not added after the physics workflow is built; it is a native mode of operating the same workflow.

The multi-encoding demonstration shows why this integration matters. Hybrid, tunnel-charge, and singlet-triplet qubits use different Hamiltonians and physical scales, yet all can be evaluated from the same reference device. The singlet-triplet example is especially illustrative: a control choice that appears favourable when viewed through charge-noise suppression alone changes when universal control is evaluated, reducing the predicted worst-gate infidelity from 0.759 to 0.0259. The specific numbers are conditional on the modelled device, but the broader capability is transferable - coupled device and qubit metrics can be explored without leaving the common design state.

The LLM integration extends this idea from interactive simulation to machine-operated design. In zenDot, the model can select calculations, modify electrical controls, edit geometry, receive quantitative solver feedback, and continue the search. The geometry experiment is particularly important because it crosses the boundary between tuning and design: changing the gate length forces the platform to reconstruct and re-solve the device rather than querying a precomputed surrogate. This is the route by which language models can become useful participants in quantum-device engineering - not by replacing physical modelling, but by navigating it.

The present agent runs are demonstrations of this operating model rather than benchmarks of optimization intelligence. They use one local language model and six evaluations per task, and the sampled trajectories are not expected to establish optimizer superiority. Likewise, the reference Si/$SiO_2$ device is an internally consistent model rather than a fit to a fabricated device, relaxation is not included in the quoted gate fidelities, and material and geometry uncertainties are not yet propagated through the full workflow. These limitations define the next validation steps without changing the platform-level result demonstrated here.

Experimental calibration, uncertainty propagation, cross-backend comparison, and larger matched-budget agent studies can all be added without changing the architecture. A calibrated device can replace the present reference state; alternative physics backends can be evaluated through the same component interface; and different agent or deterministic strategies can operate under identical design variables and objectives. zenDot therefore provides not only a toolkit for the calculations reported here, but a common substrate for increasingly automated and experimentally grounded quantum-device design.

# Conclusion

zenDot integrates semiconductor quantum-device construction, condensed-matter simulation, qubit analysis, and LLM-driven design within one quantum TCAD platform. A single material-labelled device state can be carried from geometry to experimental observables and distinct qubit models, while the same physics components can be operated through the desktop workbench, Python, or an embedded language-model agent. The Si/$SiO_2$ demonstrations show the breadth of the workflow, and the 18 agent iterations show that an LLM can act directly on the physics environment, including modifying device geometry and triggering a full re-solve. By making quantum-device simulation machine-operable without separating automation from the underlying physics, zenDot provides a foundation for autonomous, experimentally calibrated quantum-device engineering.

# Methods

## Platform architecture and component registry

A zenDot project contains a device source and a typed settings object. The device source defines material layers, gates, reservoirs, and electrical control names; the settings specify charge sectors, spin representation, solver backend, temperature, fields, noise parameters, and scan controls. Loading a project constructs the common device state consumed by registered calculation components. Each component declares a stable identifier, parameter schema, device and backend requirements, returned result type, and supported visualizations. Completed results retain component identity, parameters, units, solver metadata, and device provenance. The full component inventory and result-field definitions are provided in the Supporting Information.

## Device and electrostatics

The reference device is a lateral Si/$SiO_2$ double dot with an 8 nm Si channel, oxide, and aluminium-oxide layers. Plunger gates lie at $z = 78$ nm; source, inter-dot, and drain barrier gates lie at $z = 81$ nm. The electrostatic calculation applies material permittivities and gate boundary conditions to the scripted geometry. Localized dots are identified from the resolved channel confinement, after which lever arms, charging terms, tunnel coupling, and exchange are computed from the same device state.

*Equation 1. Material-resolved electrostatics*

$$\nabla \cdot [\epsilon(\boldsymbol{r})\nabla\phi(\boldsymbol{r})] = -\rho(\boldsymbol{r})$$

## Charge and transport

Charge-stability maps use a constant-interaction free energy evaluated over integer charge states. Sequential current is calculated with a first-order transition-rate master equation. The corresponding transport results are interpreted within this sequential-tunnelling approximation. [3,7,30]

*Equation 2. Constant-interaction free energy*

$$F(\boldsymbol{n}) = \frac{1}{2}(e\boldsymbol{n} - \boldsymbol{Q}_0)^{\mathsf{T}}\boldsymbol{C}^{-1}(e\boldsymbol{n} - \boldsymbol{Q}_0) - e\boldsymbol{n}^{\mathsf{T}}\boldsymbol{\alpha}\boldsymbol{V}$$

*Equation 3. Sequential master equation and lead current*

$$\frac{dP_i}{dt} = \sum_j \left(W_{ij}P_j - W_{ji}P_i\right), \qquad I_\ell = e\sum_{ij} s_{ij}^{(\ell)} W_{ij}^{(\ell)} P_j$$

## Encoded-qubit models

The hybrid-qubit component diagonalizes the retained three-electron logical subspace using the device-derived tunnel parameters and a declared valley splitting. The tunnel-charge component uses a two-level charge Hamiltonian with the device-derived tunnel coupling. The singlet-triplet component combines device-derived exchange $J$ with a Zeeman-energy difference $\Delta E_Z$. Each model computes its own spectrum, frequency, coherence, thermal populations, dynamics, and gate metrics from the corresponding returned arrays.

*Equation 4. Tunnel-charge Hamiltonian*

$$H_{\mathrm{CQ}} = \frac{\varepsilon}{2}\sigma_z + t_c\sigma_x, \qquad E_{01} = \sqrt{\varepsilon^2 + 4t_c^2}$$

*Equation 5. Singlet-triplet Hamiltonian*

$$H_{\mathrm{ST_0}} = \frac{J}{2}\sigma_z + \frac{\Delta E_z}{2}\sigma_x, \qquad \Omega = \sqrt{J^2 + \Delta E_z^2}$$

*Equation 6. Frequency, gate time, and quality factor*

$$f = E_{01}/h, \qquad t_g = \frac{\theta}{2\pi}\frac{h}{E_{01}}, \qquad Q = \frac{T_2^*}{t_g} = \frac{2\pi}{\theta} f T_2^*$$

## Noise, thermal state, and declared model parameters

Four parameters enter the encoded stage in addition to quantities obtained directly from the electrostatic device state: a charge-noise amplitude of 1.5 μeV $\mathrm{Hz}^{-1/2}$ at 1 Hz, a residual-gradient dephasing time of 1 μs, an electron temperature of 100 mK, and the valley splitting. zenDot evaluates the valley splitting using the 8 nm channel width and an assumed vertical channel field of 10 mV nm$^{-1}$ rather than the field returned by the Poisson solution. First- and second-order splitting sensitivities are evaluated numerically at the selected operating point. Thermal populations use the calculated energy levels and retained degeneracies. The quality factor is $Q = T_2^*/t_g$, where $t_g$ depends on the rotation angle $\theta$; the quoted hybrid value uses a π rotation, whereas the tunnel-charge and singlet-triplet values use π/2 rotations.

*Equation 7. First- and second-order splitting noise*

$$\sigma_E^2 \simeq \left(\frac{\partial E_{01}}{\partial \lambda}\right)^2 \sigma_\lambda^2 + \frac{1}{2}\left(\frac{\partial^2 E_{01}}{\partial \lambda^2}\right)^2 \sigma_\lambda^4$$

*Equation 8. Thermal occupation*

$$p_i = \frac{g_i \exp[-E_i/(k_\mathrm{B}T)]}{\sum_j g_j \exp[-E_j/(k_\mathrm{B}T)]}$$

### LLM-driven design execution

Each LLM design task specifies an objective, metric path, optimization direction, one allowed design variable, physical bounds, and a completed-simulation budget. The model returns a structured action that selects a registered component and proposes the next parameter value. zenDot verifies that the action is syntactically complete and compatible with the declared task and current device before execution. The selected component then performs the physics calculation, and the objective is read from the returned result rather than from the model text. The measured history is passed back to the design loop for subsequent proposals. Complete task declarations, prompts, replies, execution records, and pseudocode are provided in the Supporting Information.

*Equation 9. Bounded optimization over valid device states*

$$x_* = \underset{x \in \mathcal{B} \cap \mathcal{S}}{\arg\,\mathrm{best}}\; R[\mathcal{C}(D, x)]$$

## Data availability

The source data underlying all seven figures, including device, characterization, encoding, state, and optimization records, are provided with this paper as a unified Source Data package. Saved dialogue records and cross-study validation summaries are included in the Supporting Information.

## Code availability

The zenDot source code, device definitions, calculation stages, and figure-building scripts used in this study are being prepared for commercial development and are not publicly available. Non-commercial academic evaluation access may be requested from the corresponding author and may be provided under applicable confidentiality, evaluation, or licensing terms. Confidential access can be made available to editors and reviewers during peer review where required.

## Acknowledgements

The author acknowledges the use of OpenAI Codex and Anthropic Claude Code for software-development assistance, code review, documentation, and language editing. All AI-assisted outputs were reviewed, tested, and validated by the author, who takes responsibility for the final code and manuscript.

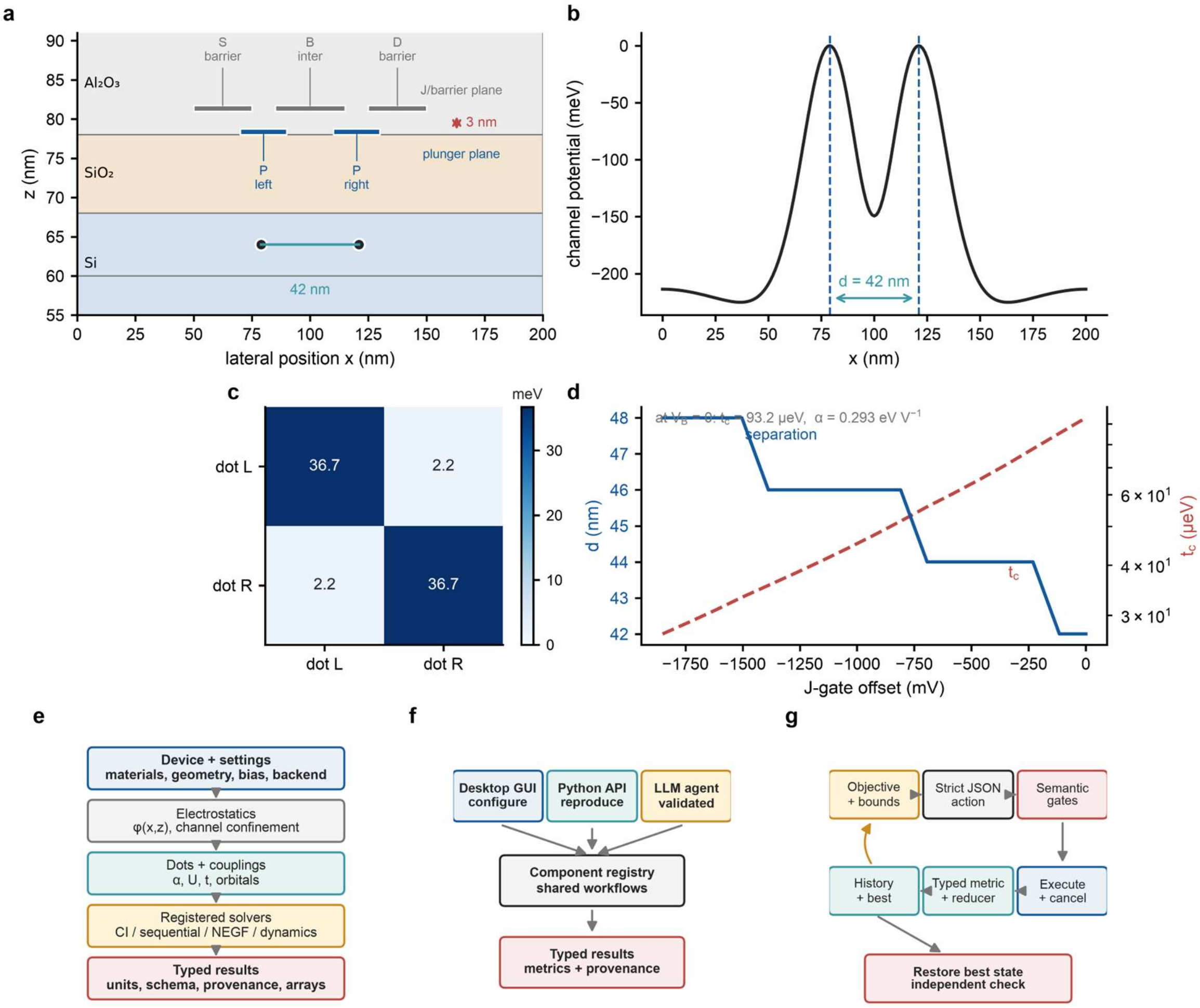

Fig. 1 | zenDot connects one device state to human and LLM-operated quantum-device simulation. a, Material-labelled Si/$SiO_2$ layer stack. b, Channel confinement returned by the electrostatic solution, resolving two minima 42 nm apart. c, Charging matrix derived from the same device state. d, Inter-dot gate control of separation and tunnel coupling. e, Device-to-result workflow shared across the platform. f, Desktop workbench, Python API, and LLM agent access the same component registry. g, Physics-aware LLM execution: proposed design actions are checked, executed by registered solvers, and returned as quantitative results for the next iteration.

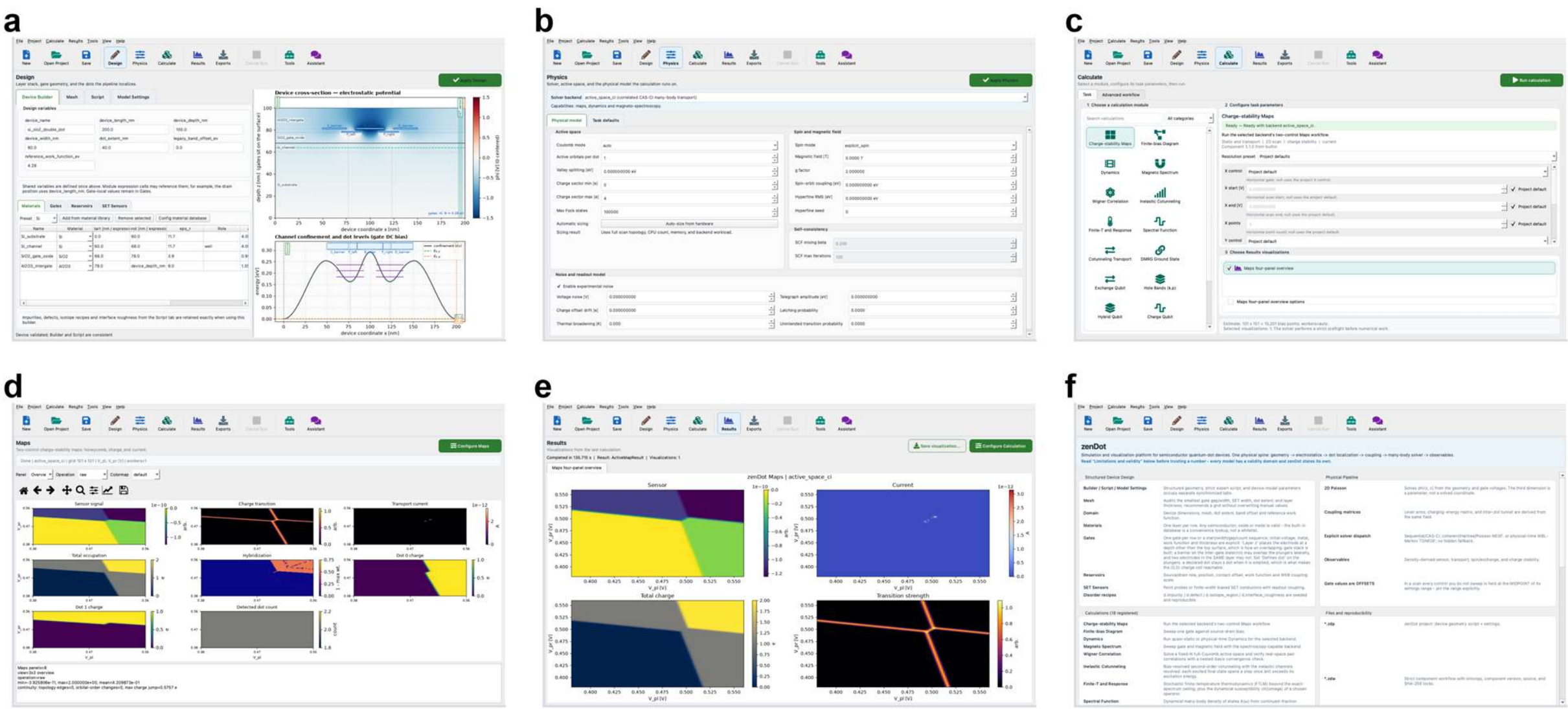

Fig. 2 | The zenDot desktop workbench. a, Device builder and derived confinement. b, Physical and solver settings. c, Registry-generated calculation components and compatible visualizations. d, Charge-stability and transport-map workspace. e, Result visualization and export. f, Integrated documentation. The same registered components are accessible through Python and the LLM Assistant.

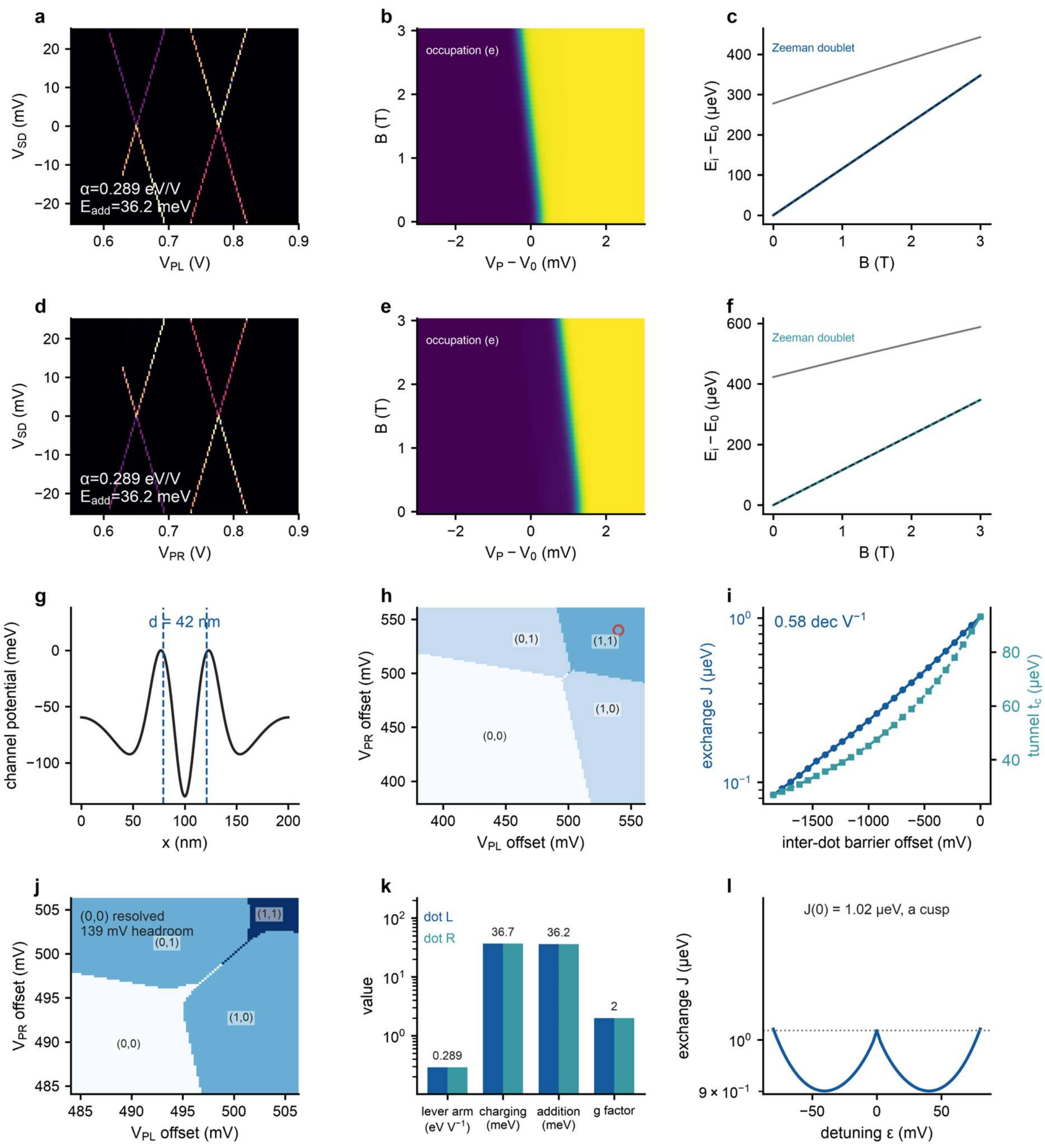

Fig. 3 | Device-level characterization generated from the reference $Si/SiO_2$ double dot. a-f, Coulomb diamonds, magnetospectroscopy, and Zeeman levels calculated independently for the left and right dots. g, Solved double-well confinement. h, Two-plunger charge-stability map with the selected (1,1) operating point. i, Barrier dependence of exchange and tunnel coupling. j, Empty-cell headroom below the first charging line. k, Left-right consistency of single-dot quantities. l, Even exchange J(ε) with a cusp-shaped local maximum at ε = 0.

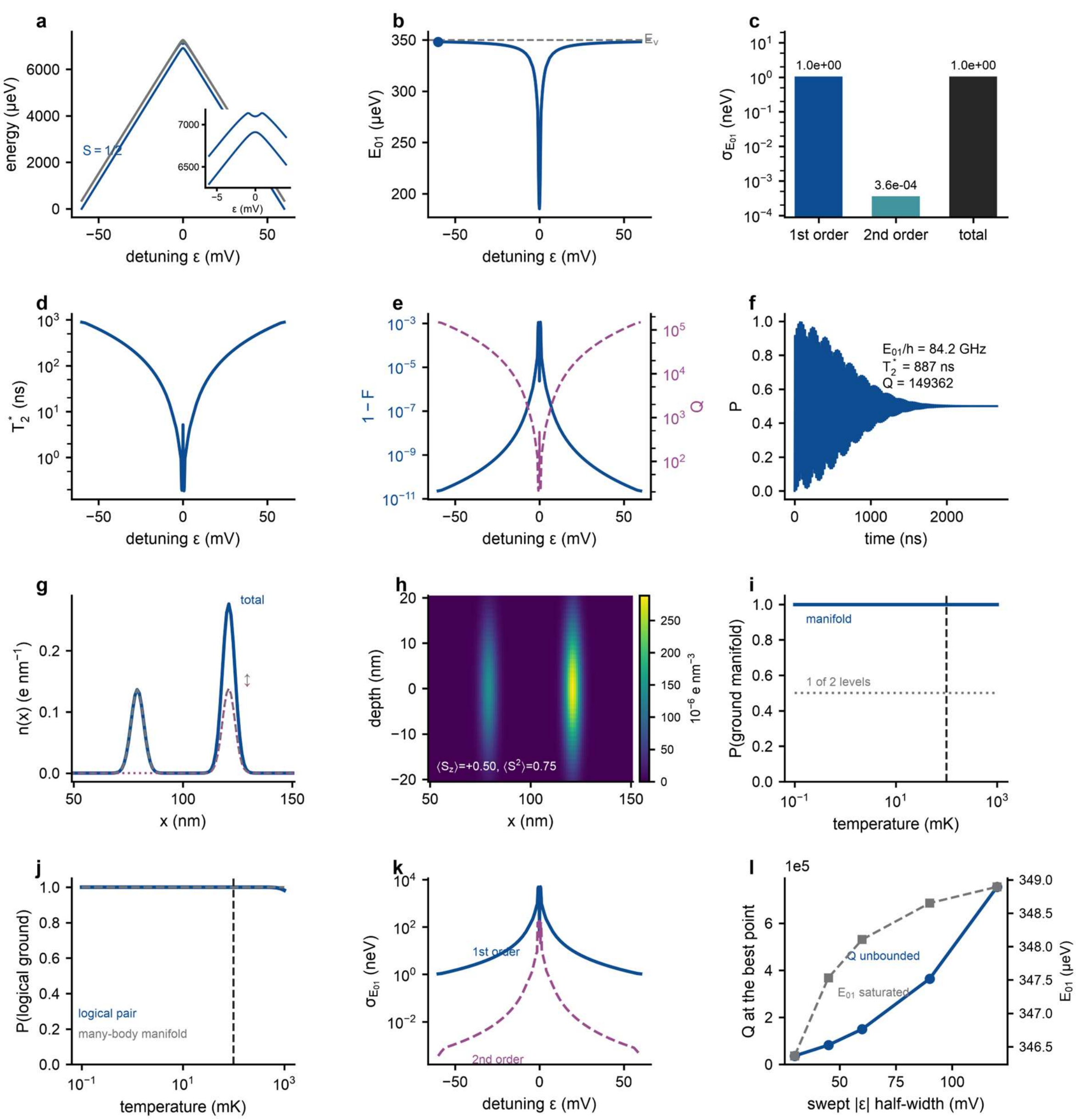

Fig. 4 | Hybrid-qubit demonstration on the calibrated device. a, Three-electron spectrum. b, Logical splitting $E_{01}$ approaching the valley splitting. c-e, Noise response, dephasing time, infidelity, and quality factor. f, Dephased logical fringe. g,h, Real-space and depth-resolved density. i,j, Thermal populations and logical initialization. k, First- and second-order amplitude channels. l, Dependence of the reported quality factor on detuning window. At the reported operating point $E_{01}$ = 348.1 μeV, the frequency is 84.2 GHz, $T_2^*$ = 887.3 ns, and the logical initialization fidelity is 1.000 at 100 mK.

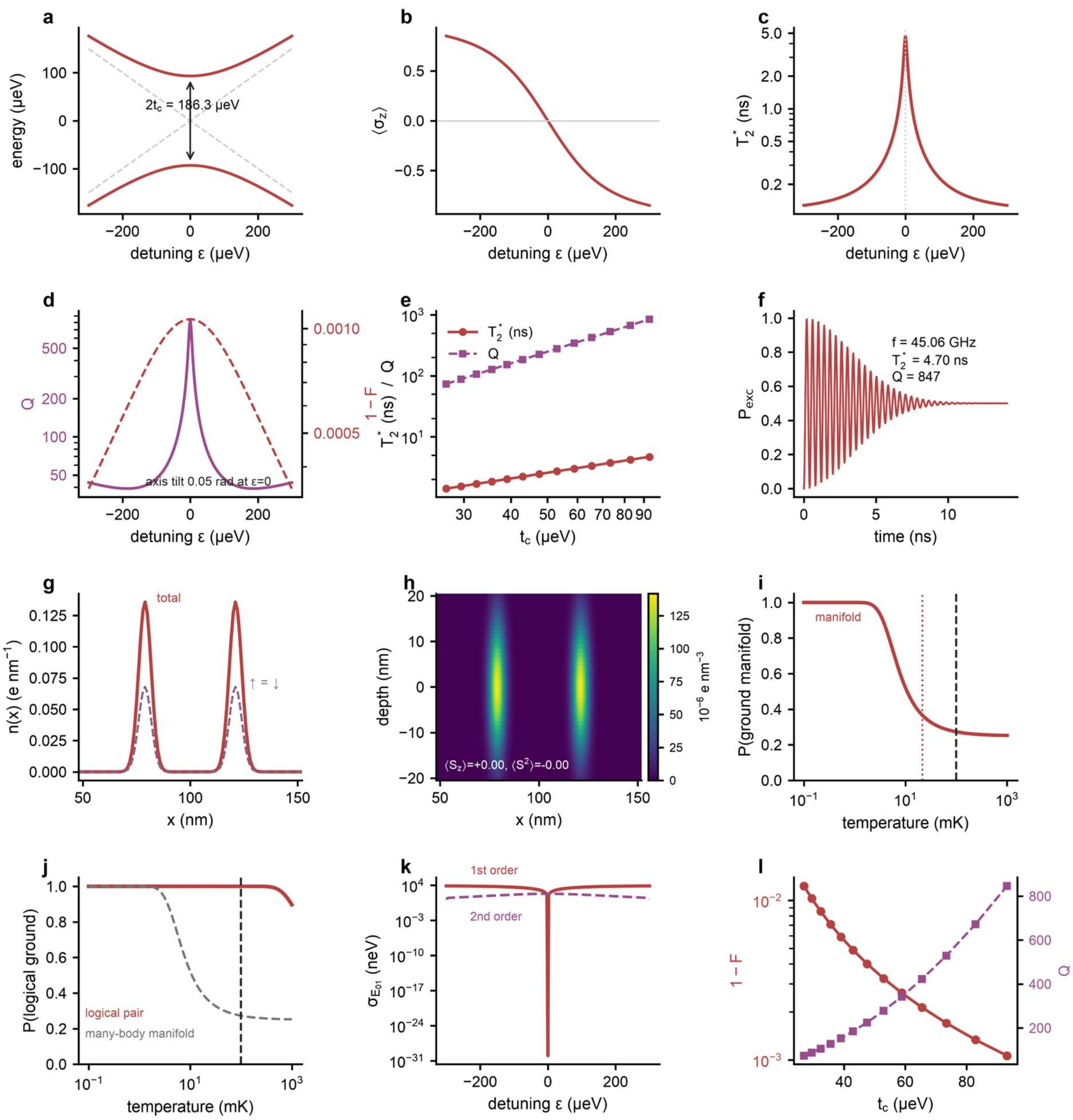

Fig. 5 | Tunnel-charge-qubit demonstration. a, Charge anticrossing with minimum splitting $2t_c$. b, Charge dipole. c,d, Dephasing and gate-error metrics across detuning. e, Dependence on tunnel coupling. f, Larmor fringe. g,h, Density. i,j, Many-body and logical thermal populations. k, First- and second-order amplitude channels. l, Barrier-controlled gate performance. At zero detuning, the frequency is 45.1 GHz, $T_2^*$ = 4.70 ns, $Q$ = 847 for a π/2 rotation, and the logical initialization fidelity is 1.000 at 100 mK.

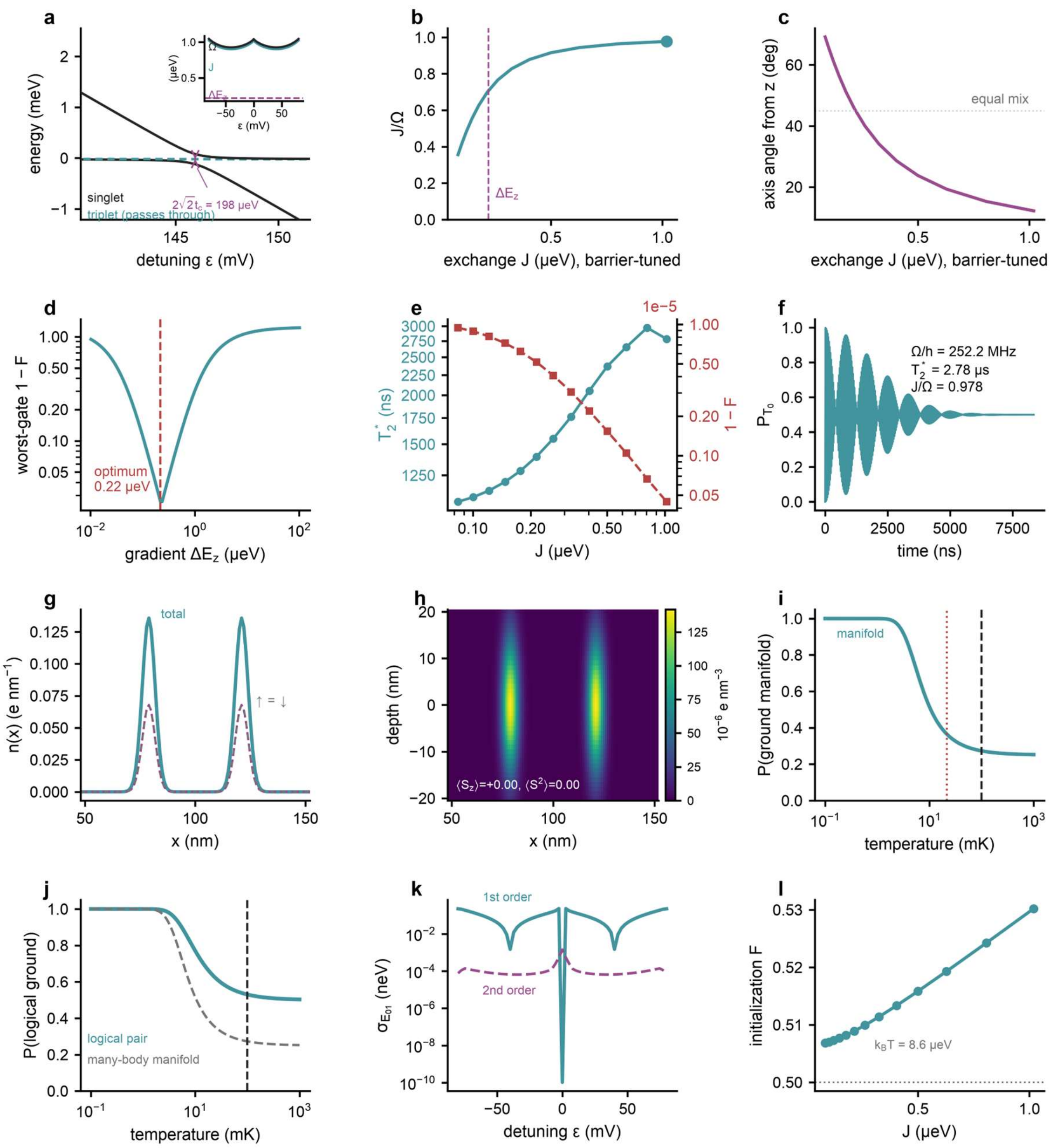

Fig. 6 | Singlet-triplet-qubit demonstration and platform-level control scan. a, Two-electron spectrum and logical splitting $\Omega = (J^2 + \Delta E_Z^2)^{1/2}$. b,c, Charge-noise suppression and rotation-axis angle. d, Worst-gate infidelity versus Zeeman gradient. e, Dephasing and gate error versus exchange. f, Dephased oscillation. g,h, Density. i,j, Thermal populations. k, Noise channels. l, Initialization versus exchange. The selected operating point gives $\Omega$ = 1.043 μeV, 252.2 MHz, $T_2^*$ = 2782.6 ns, and logical initialization fidelity 0.530 at 100 mK.

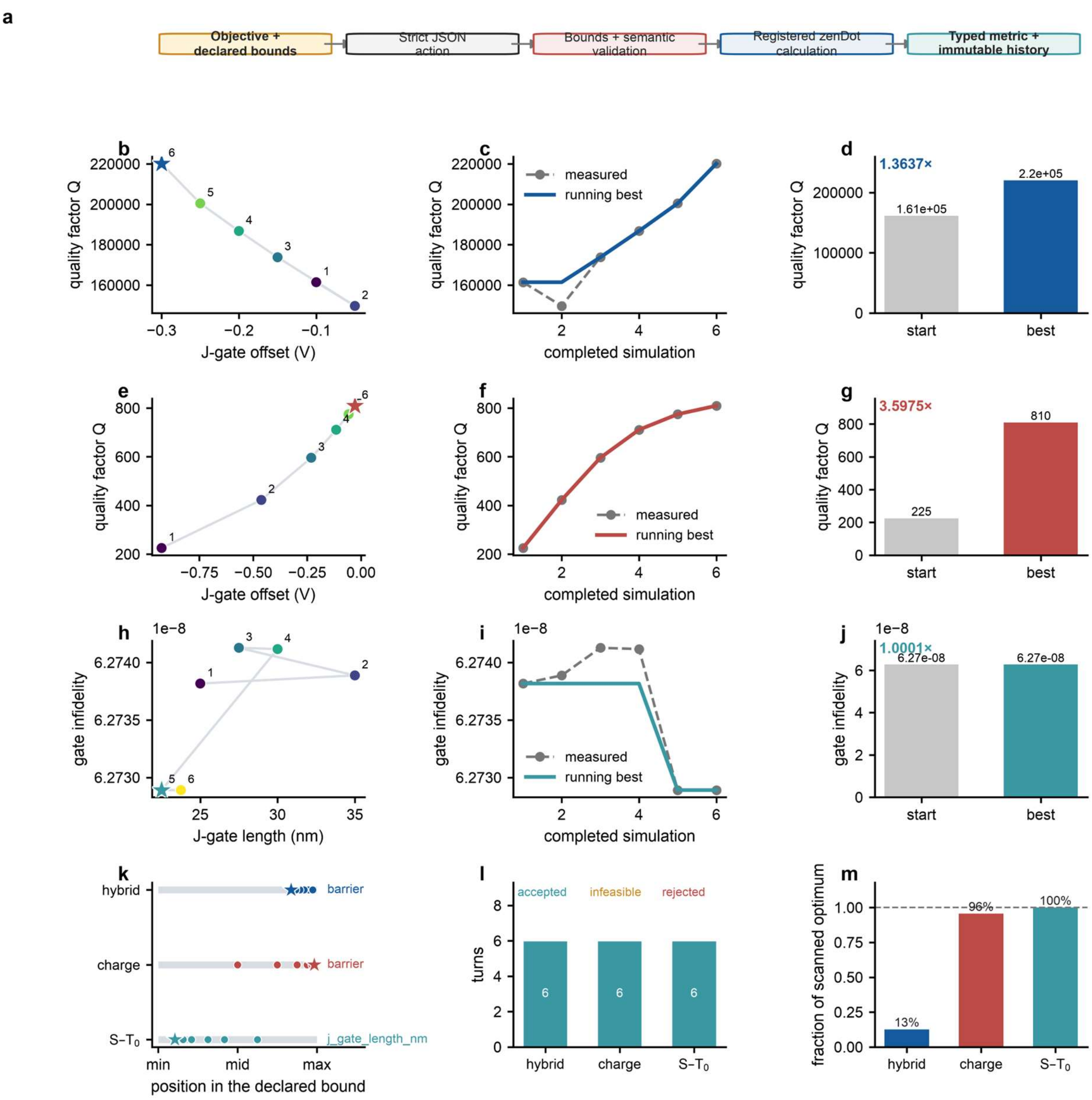

Fig. 7 | LLM-driven quantum-device design in zenDot. a, The agent receives an objective, allowed design variable, bounds, and simulation budget, then operates registered physics components and receives solver-returned metrics. b-d, Hybrid-qubit design trajectory. e-g, Tunnel-charge design trajectory. h-j, Geometry-level singlet-triplet trajectory, including device reconstruction and a full re-solve after each geometric change. k, Sampled locations within the declared design ranges. l, Execution status of the 18 completed iterations. m, Comparison with landscapes scanned using the same physics components. Full prompts, replies, numerical histories, and validation details are provided in the Supporting Information.